\documentclass[preprint,showpacs,showkeys,preprintnumbers,amsmath,amssymb]{revtex4}
\usepackage[dvips]{epsfig}
\usepackage{hyperref}
\usepackage{dcolumn}
\usepackage{bm}

\usepackage{amssymb}

\newcommand{\rem}[1]{}

\begin{document}

\title{
Effects of axial torsion on sp carbon atomic nanowires}

\author{Luca Ravagnan$^{1,2}$}
\author{Nicola Manini$^{1,3}$}
\author{Eugenio Cinquanta$^{1,2}$\footnote{Present address: Material
    Science Department - University of Milan Bicocca, Via Cozzi 53, 20125
    Milano, Italy}}
\author{Giovanni Onida$^{1,3}$}
\email{giovanni.onida@mi.infn.it}
\author{Davide Sangalli$^{1,3}$}
\author{Carlo Motta$^{1,3}$}
\author{Michele Devetta$^{1,2}$}
\author{Andrea Bordoni$^{1,2}$}
\author{Paolo Piseri$^{1,2}$}
\author{Paolo Milani$^{1,2}$}
\email{pmilani@mi.infn.it}
\affiliation{$^1$Physics Department and INFM - University of Milan, Via Celoria 16, 20133 Milano, Italy }
\affiliation{$^2$CIMAINA - Via Celoria 16, 20133 Milano, Italy }
\affiliation{$^3$European Theoretical Spectroscopy Facility, Via Celoria 16, 20133 Milano, Italy }

\date{April 22, 2009}

\begin{abstract}
{\it Ab-initio} calculations within Density Functional Theory
combined with experimental Raman spectra on cluster-beam deposited 
pure carbon films provide a consistent picture of
sp-carbon chains stabilized by sp$^3$ or sp$^2$ terminations,
the latter being sensitive to torsional strain.
This unexplored effect promises many exciting applications since it allows
one to modify the conductive states near the Fermi level and to switch 
on and off the on-chain $\pi$-electron magnetism.
\end{abstract}



\maketitle

Post-silicon electronics has seen the recent opening of entirely new
perspectives along the way of carbon-based devices.
By proper nanoscale design, entirely carbon-made transistors have been
realized \cite{Lin08}.
Future applications have been devised, including bio-nanotechnology ones
such as devices for fast DNA reading \cite{Postma08}. 
%
Even considering only 
well-demonstrated applications, the potential of
carbon-based electronics is undoubtedly enormous, as testified by the
realization of non-volatile memories based on two-terminals atomic-scale
switches \cite{Standley08} and bistable graphitic memories \cite{YLi08}.
Specifically, these structural memory effects have been explained by the
formation of carbon chains made by a few aligned sp-hybridized atoms
bridging a nanometric gap \cite{Standley08}.

In this context, the production of pure carbon nanostructured films
with coexisting sp and sp$^2$ hybridization \cite{Ravagnan02,Ravagnan07}
opens the exciting possibility to tailor
complex carbon-based nanostructures with linear chains made of
sp-hybridized C atoms connecting graphitic nano-objects.
However, despite several theoretical studies devoted to sp carbon nanowires
\cite{Lang98e00,Quian08,Crljen07,Weimer05} classified either as cumulenes
(virtually conducting, characterized by double C-C bonds) or polyynes
(large-gap insulators with alternating single and triple bonds), the
implications associated with the nanoscale geometrical manipulation of
hybrid sp+sp$^2$ carbon systems are still largely unexplored.

In this Letter we show that sp nanowires can be stabilized
effectively by termination on graphitic nanofragments, and that in the
resulting structures the $\simeq 1$~nm-long linear atomic chains can be 
{\it torsionally} stiff, due to the
broken axial symmetry with staggered 
$\pi$-bonds.
This stiffness is rich of consequences. We explore here how the structural,
vibrational and electronic properties of such chains are affected by the
nature and {\it geometry} of the termination.
In particular, we show that sp$^2$ bonding to graphitic fragments and
graphene nanoribbons (NRs) produces remarkably stable structures, with
cumulene-type chains displaying a non-negligible bond length alternation
(BLA), so that the traditional categories of polyynes
(alternating single-triple bonds, yielding a large BLA) 
and cumulenes
(double bonds, negligible BLA) 
appear too simplistic for the description of these systems.
%
%
Torsional deformations are found to affect the BLA, electronic gap,
stretching vibrational frequencies and spin magnetization of the chains.

We study these effects in realistic nanostructures,
including carbon chains bound to graphitic fragments.
We perform all calculations within Density Functional Theory in the
Local Spin Density Approximation, 
using a plane-wave basis as
implemented in the ESPRESSO \cite{espresso} suite
\footnote {
  We use the C.pz-rrkjus.UPF ultrasoft pseudopotential from
  Ref.~\cite{espresso} with a wavefunction/charge cutoff of 30/240~Ry, and
  relax all atomic positions until the largest residual force is
  $<2\times10^{-4}$~Ry/$a_0$ (8~pN).
  In the case of periodic graphene NRs, we adopt supercells with three
  hexagonal units along the periodic direction and at least 7~\AA\ of
  vacuum separating periodic images in the two other directions, optimizing
  the lattice constant until the stress tensor drops below
  $2\times10^{-5}$~Ry/$a_0^3$.
  We sample the Brillouin zone with at least 13 $k$-points in each periodic
  direction and only $k=0$ in non-periodic directions. Numerical details are
  similar to those validated and used, e.g., in Ref.\ \cite{Wassmann08}.}.

\begin{figure}
\centering
\epsfig{file=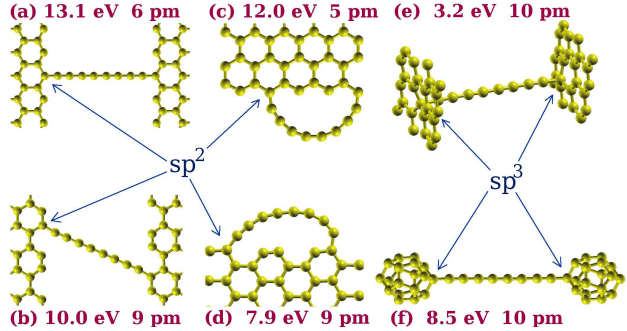,angle=0,width=8.2cm,clip=}
\caption{\label{structures} (Color online)
A few representative structures involving an 8-atoms sp-bonded carbon
chain terminated on sp$^2$ carbon fragments [(a-e): NRs; (f): C$_{20}$].
Either edge termination  [(a-d), sp$^2$-like] or termination
on an internal atom of the fragment [(e-f), sp$^3$-like] is possible.
Binding energies (with respect to the uncapped straight chain plus fully
relaxed sp$^2$ fragments) and BLA are reported.
}
\end{figure}

Figure~\ref{structures} displays a few of the studied systems involving
either sp$^2$ or sp$^3$ bonding of a sp nanowire with an sp$^2$-type
fragment. 
The chosen end-capping nanostructures include planar graphitic fragments
and closed-cages clusters (here, C$_{20}$, the most curved fullerene).
These structures are intended to represent typical interfaces present in the
nanostructured films produced by cluster-beam deposition \cite{Bogana07}.

The nature of the terminal bonding turns out to be crucial in determining
the structure and electronic properties of the wire. An sp$^2$-kind
termination produces remarkably stable cumulene-type structures (between 7.9
and 13~eV for the formation of the two new bonds), characterized by a
BLA between 5 and 9~pm \footnote{
The BLA measures the degree of dimerization and, excluding the terminal
bonds, can be defined as $
\frac 12 \left[
\sum_{j=1}^{n_e} (d_{2j-1}+ d_{n-(2j-1)})/n_e
-\sum_{j=1}^{n_o} (d_{2j} +d_{n-2j})/n_o \right]
$, 
with
$d_i=|\vec r_i-\vec r_{i+1}|$,
$n_e=(n+2)/4$, and $n_o=n/4$ (integer part).
%
}.
The computed binding energy should be compared with the energy-per-bond of
2.1~eV that we obtain for a lateral attachment of the same chain to the
ribbon edge, and with the formation energy of graphene edges
\cite{Okada08,Wassmann08}; moreover, it is much larger than the reported
binding energies of carbon chains inside nanotubes \cite{Liu03}.
Figure~\ref{structures} also shows that the mere value of the BLA does not
allow one to distinguish between carbon chains which would be traditionally
classified as cumulenic (a-d) or polyynic (e-f) according to their
terminations.


For the sake of comparison we also consider standard cumulenes and polyynes, 
in the form of isolated carbon chains stabilized by hydrogen terminations.  
%
Polyynes, which can be seen conceptually as acetylene molecules with longer
alternating triple/single-bond carbon chains (C$_n$H$_2$), have been
synthesized up to a considerable length ($n=20$) \cite{Eisler05,Pino01} in
liquid and solid matrices, and also with different stabilizing
end-groups.
Their electronic and vibrational properties as isolated species have been
characterized extensively, mainly by electronic and Raman spectroscopy
\cite{Pino01,Tabata06}.
On the other hand, cumulenes C$_{n+2}$H$_4$ can be seen as C$_n$ sp chains
terminated by CH$_2$ groups, yielding all double C=C bonds, and can be
ideally thought as ethylene molecules lengthened by extra carbon atoms.
Cumulenes are more elusive and less well characterized than polyynes, due
to their fragility.
Recently, short cumulenic chains have been synthesized in their basic forms,
butatriene and hexapentaene \cite{Gu08}.
In fact, cumulenic chains are often produced in conjunction with more
complex terminations than simple CH$_2$ units, including CPh$_2$, i.e.\ 1,1
diphenyl ethyl (DPE) groups \cite{Hino03}, which we also simulate.

As one could infer from elementary valence bond or tight-binding 
considerations, 
depending on the number $n$ of carbons being even or odd,
sp$^2$-terminated cumulenes assume a D$_{2h}$ (planar) or a D$_{2d}$
(staggered) ground-state geometry respectively, due to the alternating
orientation of the $\pi$ bonds along the chain \cite{Gu08}.
Similarly, chains bonded to sp$^2$ structures are affected by
the relative orientation of their terminations.
Indeed, a memory of the orientation of the bonds of the terminating sp$^2$
carbon propagates along the sp-hybridized chain,
so that even-$n$ chains tend to relax to a configuration where the
termination sp$^2$ planes coincide, while odd-$n$ chains tend to keep their
terminations at a twist angle $\theta=90^\circ$.
As a consequence, despite their purely one-dimensional 
nature, sp$^2$-terminated
carbon chains display a non-vanishing torsional stiffness, 
no matter if they are straight or bent as in Fig.~\ref{structures}c,d.
In contrast, ideally polyynic chains
(i.e.\ those terminated at a sp$^3$
site, with a pure single-triple bond alternation)
are almost completely
free to rotate around their axis, but suffer from an obvious frustration
when the number of atoms is odd since the long-short bond alternation must
swap at their middle \cite{Khoo08}.
Importantly, in nanostructured cluster-assembled carbon characterized by a
complex three-dimensional arrangement of graphitic fragments and sp chains
\cite{Ravagnan02,Ravagnan07}, a large number of the chains binding to
sp$^2$ structures are not free to relax their terminations to the preferred
angular geometry, and must hence be expected to be, in general, strained
torsionally.

Since the simulation of carbon-only structures such as graphene NR bridged
by chains allow us to investigate few relative angular arrangements only,
we extend our study also to chains with simpler
saturating ligands, namely CH$_2$ and DPE.
The latter turns out to reproduce better the behavior of a large
(potentially semi-infinite) graphitic fragment, which, at variance with
CH$_2$, shares with the chain only a fraction of its unsaturated p$_z$
electron, which is partly delocalized over an extended aromatic sp$^2$
structure.

\begin{figure}
\centering
\epsfig{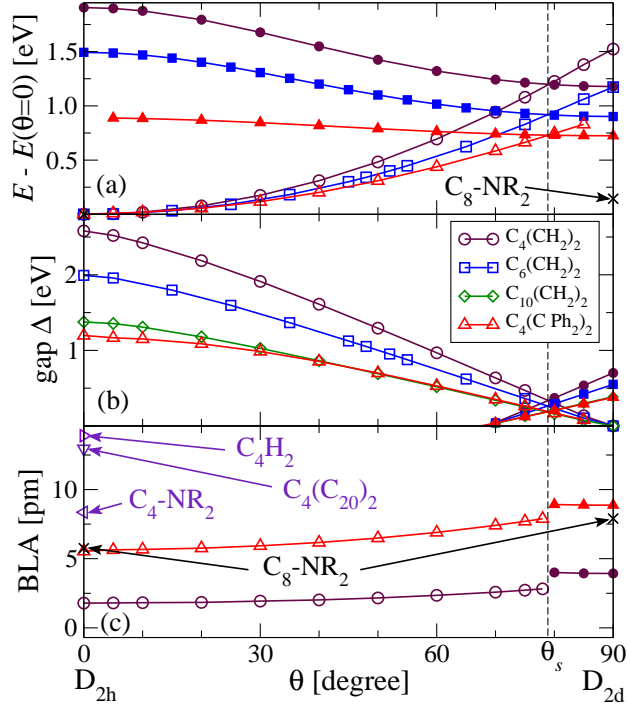}
\caption{\label{theta_dep} (Color online)
Total torsional energy (a), Kohn-Sham electronic gap (b), and bond-length alternation (c) as a function of the twist angle $\theta$ for representative
even-numbered sp-carbon chains with different terminations.
Open/filled symbols refer to the low/high-spin electronic configurations.
%
}
\end{figure}

Figure~\ref{theta_dep} summarizes the influence of different end groups on
the BLA and, for sp$^2$ termination, the torsional strain energy and the
Kohn-Sham electronic gap of even-$n$ chains as a function of $\theta$.
Interestingly, largely strained chains undergo a magnetic instability,
turning spin polarized. The reason is the quasi-degeneracy of two
$\pi$-bonding/antibonding electronic levels near the Fermi energy
illustrated by the closing of the gap, Fig.~\ref{theta_dep}b.
Remarkably, in even-numbered chains of all considered lengths, no matter if
CH$_2$- or DPE-terminated, the switching to a spin-polarized
configuration takes place at the same twist angle $\theta_s\simeq
79^\circ$, highlighted in Fig.~\ref{theta_dep} \footnote{
Odd-numbered chains show a reversed behavior, with high-spin states near
the energetically unfavorable planar geometry $\theta\simeq 0^\circ$.
}.
%
%
This $\theta_s$ invariance implies that the energy gap $\Delta$ 
and the exchange splitting of the electrons near the Fermi level scale in
the same way.

Calculations show that the BLA of the sp$^2$-terminated chains varies
substantially with the nature of the termination itself \cite{Weimer05}.
The length of the extremal bond (i.e.\ the one connecting the last sp
carbon with sp$^{2}$/sp$^{3}$-hybridized ligand), which correlates with the
BLA, is minimal in the case of a simple CH$_2$ termination, but increases
substantially in DPE-terminated chains, assumes even larger values in
NR-terminated wires, and is maximum for polyynic-type terminations, see
Fig.~\ref{structures} and Fig.~\ref{theta_dep}c.
The torsional barrier is consistently smaller for NR terminations, as
indicated by the cross at $\theta =90^\circ$ in Fig.~\ref{theta_dep}a.

In the light of the above results, chain-termination details are expected
to influence the vibrational properties as well.
The latter offer an invaluable opportunity to check if the the considered
structures are representative of those present in nanostructured
cluster-assembled films, for which Raman spectra are the main experimental
evidence of the presence of linear carbon chains.
In fact, in previous works some of us showed that the Raman fingerprint of
carbyne chains in sp-sp$^2$ carbon is characterized by a broad feature,
where 2 components C1 and C2 peaked at 1980 and 2100~cm$^{-1}$ respectively
can be recognized \cite{Ravagnan02,Ravagnan07}.
Traditionally these features were attributed generically to cumulenes
(C1) and polyynes (C2).
%
We hence calculate the phonon frequencies and eigenvectors
of
the
structures exemplified in
Figs.~\ref{structures}a and \ref{structures}f,
plus CH$_2$- and DPE-terminated carbynes of several lengths,
using standard
Density-Functional Perturbation Theory \cite{espresso}\footnote{
  All phonon calculations start by fully relaxing all degrees of freedom
  except, in the case of torsionally strained C$_n$(CH$_2$)$_2$ systems,
  the angular coordinates of the H atoms around the molecular axis.
}.
As a benchmark, our theoretical C-C streching modes of polyynes
C$_{n}$H$_2$ ($n=8-12$) match the experimental frequencies \cite{Tabata06}
to within 40~cm$^{-1}$.

Beside several bending and long-wavelength stretching modes, whose low
frequency falls in the same range as the vibrations of graphitic and
diamond-like carbon material, short linear carbon chains display a few
characteristic Raman-active stretching modes in the range $1800\div
2300$~cm$^{-1}$.
One mode, sometimes named the $\alpha$-mode in the literature
\cite{Tabata06}, shows a displacement pattern localized near the chain
center, and usually bears the strongest Raman intensity \cite{Tabata06}.
%
Since the displacements at the chain ends are less than 10\% of those of
the central atoms,
the frequency of
the $\alpha$-mode is
almost unaffected by
the mass of the termination
(e.g.\ calculations for C$_6$H$_4$ and C$_8$H$_4$ with 1000-times increased
hydrogen mass give frequency shifts of less than 1~cm$^{-1}$).

\begin{figure}
\centering
\epsfig{file=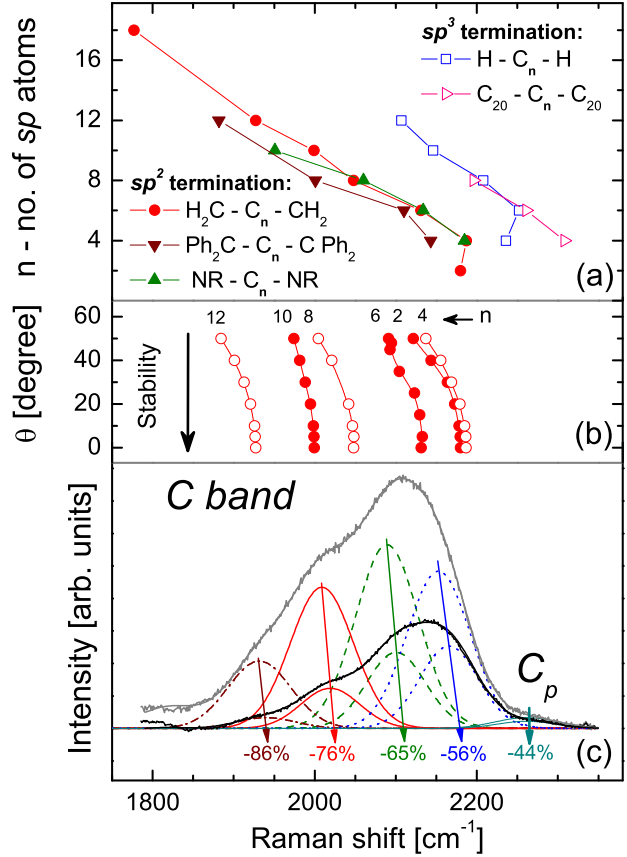,width=8.2cm,angle=0,clip=}
\caption{\label{exper} (Color online)
(a) The computed frequency of the Raman $\alpha$-mode (horizontal scale) for
  carbon chains of different length $n$ (vertical scale) and with different
  terminations
(b) The softening of this mode for CH$_2$-terminated chains as a function
   of the twist angle $\theta$ (vertical scale)
(c) The experimental Raman spectra of pristine cluster-assembled
  sp-sp$^2$ film (grey line) and of the same material after 2 days
  exposure to He, 100~Torr (black line). The underlying Gaussians
  report the empirical analysis of both spectra, resulting in 5 components
  at frequencies separated by approximately 80~cm$^{-1}$.
  The individual components display different decays, beside becoming
  narrower and undergoing a $\sim 10$~cm$^{-1}$ blueshift.
}
\end{figure}

%
The stretching frequencies of sp chains turn out to be influenced by: (i)
the type of termination (sp$^3$ vs.\ sp$^2$); (ii) the chain length, with
even/odd alternation effects; (iii) for sp$^2$ termination, the relative
orientation of the termination themselves, with effects of torsional
strain.
The calculated frequencies of the high-energy Raman-active modes display a
clear distinction between sp$^2$ and sp$^3$-terminated chains, as shown in
Fig.~\ref{exper}.
Only even-numbered chains are reported, since odd chains lack $\alpha$
modes, and have in general much smaller Raman cross sections \footnote{
 For the simplest structures (CH$_2$-terminated chains) we also  compute
 Raman intensities.
}.

Figure~\ref{exper}c displays the {\it in-situ} Raman spectrum of an
sp-sp$^2$ nanostructured-carbon (ns-C) film \cite{Ravagnan02} in the
carbyne region, measured using the 488~nm line of an Ar$^+$ laser and
acquired with very high signal-to-noise ratio.
The spectrum of the as-deposited material is compared to that obtained
after exposure of the film to He in order to promote sp chain decay
\cite{Casari04}.
Clearly, a description in terms of two peaks only cannot account for the
complex structure and decay observed.
In particular, the C$_{p}$ component at the highest frequency (peaked at
2260~cm$^{-1}$) can be attributed uniquely to short polyynic chains, as it
is higher than any cumulenic $\alpha$ mode (see Fig.~\ref{exper}a), while
the other components can be related both to polyynes and cumulenes of
different chain length.
As illustrated in Fig.~\ref{exper}c, after He exposal, individual
components have different evolutions during the C band decay, and in
particular the peaks at lower energy, corresponding to longer chains,
decay faster than the higher-energy ones (i.e. shorter chains).
Furthermore, the C$_{p}$ peak does not shift during the decay nor change its
width, while all lower peaks are blue-shifted by $\sim 10$~cm$^{-1}$ and
narrowed by $\sim 7$~cm$^{-1}$.
Indeed calculations, summarized in Fig.~\ref{exper}b, show that the
high-frequency stretching modes of torsionally-strained CH$_2$-terminated
chains are affected quite strongly by the twist angle, with a redshift up
to $\sim 100$~cm$^{-1}$. However, since chains with smaller torsional
barrier (such as those bound to DPE and nanoribbons) show smaller
redshifts, this effect evaluated for CH$_2$-terminated chains should be
considered as an upper limit for realistic pure-carbon nanostructures.

The observed blue shift of the peaks accompanying the decay
can then be explained if each peak is
related to a particular family of cumulenes, having all the same length but
different strain: the more strained chains, having softer Raman
modes, decay faster than the others, resulting in a net blue shift and
narrowing of the peak.
A faster decay of torsionally strained vibrationally red-shifted
cumulene-type is indeed to be expected due to their higher total energy
(Fig.~\ref{theta_dep}a).
On the contrary, no torsional strain applies to polyynes, and this is why
the C$_{p}$ peak does not shift.


In summary, we performed {\it ab-initio} total-energy and phonon
 calculations on a selected range of model structures sampling
 significantly the infinite variety of three-dimensional arrangements
 of linear carbon chains bridging graphitic fragments in different
 hybridization states. Theoretical results suggest that sp-carbon
 chains are stabilized in particular by bonding to the edges of graphitic
 nanofragments, and allow us to interpret the
 nontrivial features and decay of experimental Raman
 spectra of cluster-beam deposited pure carbon films.
 Moreover, the data for sp$^2$-terminated chains point towards a
 rich phenomenology driven by even/odd alternation effects and
 by the effects of torsional strain.
 The latter modifies the electronic states near the Fermi level,
suggesting
 the possibility to control the nanowire conductance\cite{Khoo08}, 
 optical properties, and
 spin magnetization, {\em purely by twisting} its sp$^2$ termination,
 e.g.\ by coupling terminating graphene sheets with micromachined
 torsional actuators \cite{Yeh99}.
 Linear carbon chains bridging graphene nanogaps, recently
 proposed
 as an explanation of the conductance switching 
 in two-terminals graphene devices \cite{Standley08,YLi08},
 could hence acquire an important role in future
 carbon-based electronics.


We acknowledge support
by the Italian MIUR through PRIN-2006025747
and by the European Union through
the ETSF-I3 project (Grant Agreement No.\ 211956 / ETSF User Project No.\ 62).


\end{document}